\documentclass[twocolumn,showpacs,preprintnumbers,amsmath,amssymb,
superscriptaddress]
{revtex4}
\usepackage{graphics}
\usepackage{epsfig}
\usepackage{dcolumn}
\usepackage{bm}
\newcommand{\nn}{\nonumber}
\newcommand{\lb}{\label}
\newcommand{\be}{\begin{eqnarray}}
\newcommand{\ee}{\end{eqnarray}}

\renewcommand{\d}{\mbox{${\rm d}$}}

\begin{document}
\title{Gauss-Bonnet Brane Cosmology with Radion Stabilization}
\author{G.L.~Alberghi}
\email{alberghi@bo.infn.it}
\affiliation{Department of Physics, University of Bologna,
Via Irnerio~46, 40126~Bologna, Italy.}
\affiliation{Department of Astronomy, University of Bologna, Italy.}
\affiliation{I.N.F.N., Sezione di Bologna, Italy.}
\author{A.~Tronconi}
\email{tronconi@bo.infn.it}
\affiliation{Department of Physics, University of Bologna,
Via Irnerio~46, 40126~Bologna, Italy.}
\affiliation{I.N.F.N., Sezione di Bologna, Italy.}
\begin{abstract}
We study cosmology in a five-dimensional brane-world with a stabilizing
effective potential for 
the radion and matter localized on the two branes.
We consider the corrections induced by the Gauss-Bonnet contribution to the
total action  
performing an expansion around the two possible static solutions
up to second order in the ratio between brane matter energy density 
and brane tensions. The Friedmann and acceleration equations on the visible brane are 
obtained and discussed. 
\end{abstract}
\maketitle
\section{Introduction}
\label{I}
Following the approach used in \cite{Alberghi:2005ah} we analyze the
cosmological implications of the 
GB action in a RS setup
(\cite{Randall:1999ee,Randall:1999vf,Goldberger:1999uk,Nojiri:2002hz,Kim})with 
radion stabilization through 
the addition of an effective term for the radion field to the total action. 
The Einstein equations are solved,
assuming the adiabatic evolution of the fluids on the branes, up to
the second order in $\rho/V$ where $\rho$ is the energy density
of the fluids and $V$ the tension of the branes. Finally the corrections
to the Friedmann and acceleration equations are analyzed and compared to
the case in which the GB term is switched off.\\
%
%
\section{Einstein equations and Static Solutions}
\label{II}
Let us start with the following action for the five-dimensional bulk dynamics
\be
&&S_{bulk}=\frac{1}{k^2}\int d^5 x\sqrt{-g}\left(\mathcal{L}_{EH}+\frac{\alpha}{2}\mathcal{L}_{GB}\right)
\ee
where
\be
\mathcal{L}_{EH}&=&R+2k^2(\Lambda-U)
\\
\nn\\
\mathcal{L}_{GB}&=&R^2-4R^{AB}R_{AB}+R^{ABCD}R_{ABCD}
\ee
with $U$ the effective contribution which stabilizes the size of the
extra-dimension (\cite{Csaki:1999mp,Cline:2000tx}).
The boundary conditions for the above action are fixed by the brane contributions to 
the stress-energy tensor; in order to recover RS~I when the 
GB contribution is switched off and no matter is present on the boundaries
we consider two branes with different tensions located at 
the fifth dimension orbifold fixed points.
If we choose the following ansatz for the metric
\be\label{metric}
\d s^2&\equiv& g_{AB}\d x^A\d x^b \nn\\
&=&-n^2(y,t)\d t^2+a^2(y,t) \d x^i\d x^i+b^2(y,t)\d y^2
\ee
we can express the stabilizing contribution as
\be
U~=~M(b(y,t)-r)^2 
\ee
where $r$ is the expectation value of the radion field and the boundary terms 
are given by
\be
T_{i\;B}^A=\frac{\delta(y-y_i)}{b}
{\rm diag} (V_i+\rho_i,V_i-p_i,V_i-p_i,V_i-p_i)
\ee
where $i=p,n$, $y_i=0,1/2$ are the positions of the branes and $V_i$'s are the brane tensions.\\
By varying the total action one obtains the Einstein equations 
\be\label{Eeqs}
G_{AB}+\alpha H_{AB}&=&k^2\left(\Lambda g_{AB}+\tilde T_{AB}+\sum_i (T_{i})_{AB}\right)
\ee
where $H_{AB}$ is the second order Lovelock tensor
and $\tilde T^A_{B}=-{\rm diag}(U,U,U,U,U+b dU/db)$ is the radion
potential contribution. 
The equations (\ref{Eeqs}), given the ansatz (\ref{metric}), form
a system of four differential equations. Due to the fact $G_{AB}$ and 
$H_{AB}$ satisfy the Bianchi and Bach-Lanczos identities, respectively
only three equations are independent.
The bulk dynamics is thus determined by the system 
\be
\left\{
\begin{array}{l}
G_{tt}+\alpha H_{tt}=k^2\left(\Lambda g_{tt}+\tilde T_{tt}\right)
\\
\\
b'=(r-b)\,\left(\frac{n'}{n}
+3\frac{a'}{a}+2\frac{b'}{b}\right)
\\
\\
G_{yy}+\alpha H_{yy}=k^2\left(\Lambda g_{yy}+\tilde T_{yy}\right)\ .
\end{array}
\right.
\label{Ein1}
\ee
(see \cite{Binetruy:1999ut,Binetruy:1999hy})where a prime denotes a derivative with respect to $y$.
The boundary equations give the
following junction conditions on the two branes:
\begin{widetext}
\be\label{juncon}
\left\{
\begin{array}{l}
\displaystyle
\lim_{y\rightarrow y_i^+}
\left[\frac{4\alpha}{b^4}\left(\frac{a'}{a}\right)^3-\frac{6}{b^2}\left(1+2\alpha\frac{\dot a^2}{a^2 n^2}\right)
\frac{a'}{a}\right]=\left.\frac{k^2}{b}\left(V_i+\rho_i\right)\right|_{y=y_i}
\\
\\
\displaystyle
\lim_{y\rightarrow y_i^+}\left[
\frac{4}{b^2}\left(1+\frac{8\alpha\dot b^2}{b^2n^2}-\frac{2\alpha\dot a\dot n}{an^3}+\frac{4\alpha\dot b\dot n}{b n^3}
+\frac{2\alpha\ddot a}{an^2}-\frac{4\alpha\ddot b}{bn^2}\right)\frac{a'}{a}\right.
\left.
+\frac{2}{b^2}\left(1+\frac{2\alpha\dot a^2}{a^2n^2}
+\frac{8\alpha\dot a \dot b}{a bn^2}+\frac{16\alpha\dot a \dot n}{a n^3}
-\frac{8\alpha\ddot a}{a n^2}\right)\frac{n'}{n} \right.
\\
\\
\displaystyle
\quad \quad \quad
\left. 
-\frac{4\alpha}{b^4}\left(\frac{a'}{a}\right)^2
\frac{n'}{n}\right.
\left.-\frac{16\alpha}{b^2n^2}\left(2 \frac{\dot b}{b}+\frac{\dot n}{n}\right)
\frac{\dot a'}{a}-\frac{16\alpha}{b^2n^2}\left(\frac{\dot a}{a}\frac{\dot n'}{n}
-\frac{\ddot a'}{a}\right)\right]=\left.
-\frac{k^2}{b}\left(V_i-p_i\right)\right|_{y=y_i}
\ ,
\end{array}
\right.
\ee
\end{widetext}
where a dot denotes a derivative with respect to the universal time $t$.
Note that the eqs. (\ref{Ein1}-\ref{juncon}) 
reduce to the standard case
(without GB contribution) in the limit $\alpha\rightarrow 0$.
%
%
%
%
\label{III}
Before investigating cosmology, in the setup just described, for $\rho_i\neq 0$,
we should consider the case $\rho_i=0$ in order to find the static solutions to perturb about.
It is easy to verify that a warped metric still satisfies eqs.(\ref{Ein1}-\ref{juncon}). 
In fact, if one makes the ansatz
\be\label{staticsol}
a(y,t)=n(y,t)=\exp\left[-m r y\right];\quad b(y,t)=r
\ee
the Einstein equations and junction conditions are verified provided
\be\lb{mval}
m=\pm\sqrt{\frac{1}{2\alpha}\left(1\pm\sqrt{1-\frac{2}{3}k^2\alpha\Lambda}\right)}
\ee
with $\frac{2}{3}k^2\alpha\Lambda\le 1$ and
\be\lb{tensions}
V_p=-V_n=\frac{6m}{k^2}\left(1-\frac{2}{3}m^2\alpha\right).
\ee 
Apart from the two expected solutions which reduce to the static RS when
$\alpha\rightarrow0$, namely
\be\lb{limitmold}
m_{old}=\pm\sqrt{\frac{1}{2\alpha}\left(1-\sqrt{1-\frac{2}{3}k^2\alpha\Lambda}\right)}
\stackrel{\alpha\to 0}\longrightarrow\pm k\sqrt{\frac{\Lambda}{6}},
\ee
two additional solutions are obtained (see \cite{Deruelle:2000ge}) 
which are interestingly less sensitive to the bulk cosmological constant, 
since in the limit $ \alpha \rightarrow 0 $ 
the warp factor becomes independent of the bulk content
\be\lb{limitmnew}
m_{new}=\pm\sqrt{\frac{1}{2\alpha}\left(1+\sqrt{1-\frac{2}{3}k^2\alpha\Lambda}\right)}
\stackrel{\alpha\to 0}\longrightarrow\pm\frac{1}{\sqrt{\alpha}} 
\ee
Furthermore, due to the GB terms, warped solutions are still present when the bulk 
is filled with a positive cosmological constant as (\ref{mval}) is still real in that case.
Note that a tuning of brane tension is required only if one looks for static solutions: 
different tunings could be, as usual, treated as a perturbation $\rho_i=\Delta V_i$
and $p_i=-\Delta V_i$ and generate an expanding (contracting) phase which 
can be studied by means of the formalism we introduce in the next section.
\section{The perturbed solutions}

\label{IV}
We now investigate the cosmological evolution in the brane-world with GB contribution. 
The calculations are based on the perturbative approach ($\rho_i/V_i\ll 1$) discussed in details 
in \cite{Alberghi:2005ah} (starting from a slightly different set of equations).
The starting point is a static solution of the form (\ref{staticsol}) where 
\footnote{without loss of generality we choose the overall sign to be positive and 
consequently tune the brane at $y=0$ with a positive tension and the brane at $y=1/2$ with an opposite tension;} 
$m=|m_{old}|$ or $m=|m_{new}|$. 
When some kind of matter is added to the branes the system is detuned and evolves in time. 
The solution becomes 
\begin{subequations}
\be
&&
\!\!\!\!\!
n(y,t)=\exp\left(-m\,r\,|y|\right)\,
\left[1+\delta f_n(y,t)\right]
\label{exactansatz_n}
\\
\nonumber
\\
&&
\!\!\!\!\!
a(y,t)=a_h(t)\,\exp\left(-m\,r\,|y|\right)\,
\left[1+\delta f_a(y,t)\right]
\label{exactansatz_a}
\\
\nonumber
\\
&&
\!\!\!\!\!
b(y,t)=r+\delta f_b(y,t)
\ ,
\label{exactansatz_b}
\ee
\\
\end{subequations}
and the differential equations (\ref{Ein1}) can be written in terms 
of $H_h\equiv \dot a_h/a_h$, $\delta f_\beta$ (with $\beta=a,n,b$) and their derivatives.
Moreover we note that when such a detuning is small, one can rely on the perturbative ansatz
\begin{subequations}
\be
\delta f_\beta&\simeq&
\epsilon\,
\left[f_{\beta,p}^{(1)}(y)\,\rho_p
+f_{\beta,n}^{(1)}(y)\,\rho_n\right]
\label{genexp}
\\
&&
+\epsilon^2\,
\left[f_{\beta,p}^{(2)}(y)\,\rho_p^2
+f_{\beta,n}^{(2)}(y)\,\rho_n^2+f_ 
{\beta,m}^{(2)}(y)\,\rho_p\,\rho_n\right]
\nonumber
\\
H_h^2
&\simeq&
\epsilon\,\left(h_{h,p}^{(1)}\,\rho_p
+h_{h,n}^{(1)}\,\rho_n\right)
\nonumber
\\
&&
+\epsilon^2\,
\left(h_{h,p}^{(2)}\,\rho_{p}^2
+h_{h,n}^{(2)}\,\rho_n^2
+h_{h,m}^{(2)}\,\rho_p\,\rho_n\right)
\label{Hhom}
\ee
\end{subequations} 
and expand these equations up to second order in $\epsilon$.
Note that the time dependence in the approximate solution above is 
encoded in $\rho_i(t)$ which evolves, as usual, satisfying the constraint of the 
continuity equation for a 4-dimensional fluid with equation of state
$p_i=w_i \rho_i$ that is 
$\dot \rho_i=-3 [\dot a(y_i,t) / a(y_i,t)] (1+w_i) \rho_i. $
One can finally, order by order in $\epsilon$, solve eqs. (\ref{Ein1}),
which now contain just derivatives 
with respect to $y$, and determine the integration constants and the unknowns parameters 
in (\ref{Hhom}) by making use of the boundary conditions (\ref{juncon}).
%
Iterating the procedure described above up to second order is almost straightforward 
(see \cite{Alberghi:2005ah} for details). 
Due to the fact that the $\rho_i$'s evolve independently, one needs to solve 
15 differential equations for the coefficients $f_{\beta,i}^{(1)}$, $f_{\beta,i}^{(2)}$. 
Once such bulk equations are solved, one is left with 25 coefficient to be fixed 
(10 of which derive from first order calculations) and the junctions conditions still to be imposed. 
It is possible to divide these coefficients into two categories: 
the ones which are related to the gauge freedom of the metric 
(they have to do with the definition of the time and of the three-dimensional scale factor $a(t)$),  
and the ones that are related to the boundary dynamics. 
The junction conditions form a system of 15 independent equations which determine just the dynamical coefficients.
Five gauge coefficients can be fixed by arbitrarily defining a time evolution parameter 
and five (those related to the scale factor) may remain arbitrary since they are not present in 
the expressions of physical observables. \\
The Friedmann and the acceleration equations on the negative tension brane 
(the negative tension brane, in RS I setup with $m>0$, 
is supposed to be the 4-dimensional space-time manifold in which we live)
can thus be obtained by fixing $n(1/2,t)=1$ and expressing
$H^2\equiv(\dot a(1/2,t)/a(1/2,t))^2$ and $\ddot a/a\equiv\ddot a(1/2,t)/a(1/2,t)$ 
as functions of the time measured on the same brane.
This leads to
(we let $\epsilon\rightarrow 1$ at the end of the calculations)
\begin{widetext}
\be\label{acceleq}
\frac{\ddot a}{a}&=&-\frac{k^2m\left[(3w_p+1){\rm e}^{2mr}\rho_p+(3w_n+1)\rho_n\right]}
{6\left({\rm e}^{mr}-1\right)\left(2m^2\alpha+1\right)}+\frac{27k^2m^2\left({\rm e}^{mr}+1\right)
\left(w_n+1\right)\left(w_n+\frac{2}{3}\right)\left(w_n-\frac{1}{3}\right)}
{32Mr^2\left({\rm e}^{mr}-1\right)^2\left(2m^2\alpha+1\right)}\rho_n^2\nn\\
&&
+\left\{\frac{k^4}{144}(3w_n+1)(3w_p+1)+\frac{k^4m^2\alpha\left({\rm e}^{mr}+3\right)
\left(3w_p+1\right)\left(3w_n+1\right)}{72\left({\rm e}^{mr}-1\right)}-\frac{m^2k^2
\left(1-4m^4\alpha^2\right)}{96r^2M\left({\rm e}^{m\,r}-1\right)^2}\right.\nn\\
\phantom{\frac{A}{B}}&&\left.\times
\left[{\rm e}^{2\,m\,r}(3w_n-1)(18w_n^2+9w_pw_n+15w_p+39w_n+23)+9{\rm e}^{m\,r}
\left(6w_p^3+6w_n^3+3w_p^2w_n
\right.
\right.
\right.
\nonumber
\\
&&
\left.
\left.
\left.
\!+3w_pw_n^2+11w_p^2+11w_n^2+2w_pw_n+w_p+w_n-4\right)
+54w_p^3+27w_p^2w_n+99w_p^2-18w_pw_n\right.\right.\nn\\
&&\left.\left.+24w_p-21w_n-13\right]\phantom{\frac{A}{B}}\!\!\!\!\!\!
\right\}\frac{
{\rm e}^{m\,r}\,\rho_p\,\rho_n}{\left(2m^2\alpha-1\right)\left(2m^2\alpha+1\right)^2}-
\left\{\frac{m^2k^2(3w_p-1)\left(1-4m^4\alpha^2\right)\left({\rm e}^{m\,r}+1\right)}
{96r^2M\left({\rm e}^{m\,r}-1\right)^2}\right.\nn\\
\nonumber
\\
&&
\left.\times\left[{\rm e}^{m\,r}(27w_p^2+54w_p+23)
-9w_p-5\right]+\frac{k^4}{144}(3w_p+1)^2-\frac{k^4m^2\alpha\left(1+3{\rm e}^{mr}\right)
\left(3w_p+1\right)^2}{72\left({\rm e}^{mr}-1\right)}\right\}\nn\\
&&\times\frac{
{\rm e}^{2\,m\,r}
\rho_p^2}{\left(2m^2\alpha-1\right)\left(2m^2\alpha+1\right)^2}.
\label{2ordnegAccEq}
\ee
\be\label{friedeq}
H^2&=&\frac{k^2 m}{3\left({\rm e}^{mr}-1\right)\left(2m^2\alpha+1\right)}
\left({\rm e}^{2mr}\rho_p+\rho_n\right)-\frac{3k^2 m^2\left(1+{\rm e}^{mr}\right)
\left(w_n+1\right)\left(w_n-\frac{1}{3}\right)}{32r^2M\left({\rm e}^{mr}-1\right)^2
\left(2m^2\alpha+1\right)}\rho_n^2+\left\{\frac{k^4m^2\alpha\left({\rm e}^{mr}
+3\right)}{18\left({\rm e}^{mr}-1\right)}\right.\nn\\
&&
\left.+\frac{k^4}{36}+\frac{m^2k^2\left(1-4m^4\alpha^2\right)\left({\rm e}^{mr}+1\right)}
{48r^2M\left({\rm e}^{mr}-1\right)^2}\,
\left[18w_p^2-9w_p(w_n-1)-9w_n-1+2{\rm e}^{mr}
(9w_n^2+9w_n-4)\right]\right\}
\nonumber
\\
&&\times \frac{
{\rm e}^{mr}\,\rho_p\,\rho_n}{\left(2m^2\alpha-1\right)\left(2m^2\alpha+1\right)^2}
-\left\{\frac{m^2k^2(3w_p-1)\left(1-4m^4\alpha^2\right)
\left({\rm e}^{mr}+1\right)}
{96r^2M\left({\rm e}^{mr}-1\right)^2}\,
\left[3w_p+7-4{\rm e}^{mr}(3w_p+4)\right]\right.\nn\\
&&\left.+\frac{k^4}{36}-\frac{k^4m^2\alpha\left(3{\rm e}^{mr}+1\right)}{18\left({\rm e}^{mr}-1\right)}\right\}\frac{
{\rm e}^{2mr}\,
\rho_p^2}{\left(2m^2\alpha-1\right)\left(2m^2\alpha+1\right)^2},\nn\\
\ee
\end{widetext}
The above expressions are quite involved as they contain contributes from the 
fluids on both branes and 
reduce to the ones already found in \cite{vinet,Alberghi:2005ah} for $\alpha\rightarrow 0$ and $m=m_{old}$.  
Note that, due to the choice of the time parameter, the coefficient of $\rho_n^2$   
vanishes when $\rho_n$ behaves as vacuum energy or radiation. 
As a consequence, when $\rho_p$ is negligible, the second order terms, 
which would be otherwise responsible for a deviation from the standard cosmological evolution, 
vanish for $w_n=-1$ or $w_n=1/3$. 
This important feature of brane-world scenarios with radion 
stabilization was already present in the case $\alpha=0$ and is furthermore 
conserved when GB contribution is present.
After some algebra one can partially absorb the 
GB coupling in (\ref{friedeq})-(\ref{acceleq}) by redefining 
$\tilde k^2\equiv k^2 / (2m^2\alpha+1) $
which means that GB corrections can be observed only through the indirect contribution of positive tension brane matter. 
Furthermore note that $\alpha$ is always multiplied by $m^2$ and
consequently the GB contributions vanish in the 
limit $\alpha\rightarrow 0$ only if we consider
the case $m=m_{old}$.
In the other case $m=m_{new}$, one has 
\be
\lim_{\alpha\rightarrow 0}\alpha\cdot  m_{new}^2=1
\ee
and the evolution obeys a dynamics which is modified with respect to the standard RS case. 
In this case, keeping just the leading contributions and letting
$w_p\rightarrow -1$ 
(small brane detuning), one is left with
\be\label{friedeqalpha}
H^2&=&\frac{k^2}{9}\left[m\tilde\rho_p-\left(3w_n+4\right)\left(3w_n-1\right) \phantom{{5 \over 4}}\right.\nn\\
&&\left.\times\frac{m^2\tilde\rho_p\rho_n}{8r^2M}+
\frac{m^2\tilde\rho_p^2}{2r^2M}\right]
\ee
\be\label{acceleqalpha}
\frac{\ddot a}{a}&=&\frac{k^2}{18}\left[2m\tilde\rho_p+\left(3w_n-1\right)
\left(3w_n+4\right) \phantom {{5 \over 4}}\right.\nn\\
&&\left.\times\left(3w_n+1\right)\frac{m^2\tilde\rho_p\rho_n}{8r^2M}
+\frac{m^2\tilde\rho_p^2}{r^2M}\right]
\ee
where $\tilde\rho_p\equiv{\rm e}^{mr}\rho_p$ and one should keep 
$m \cdot\tilde\rho_p$ and $m \cdot \rho_n$ small. 
Apart from the contribution of $\tilde \rho_p$ to the 
effective cosmological constant, the term proportional to  $\rho_n ^2$ is negligible 
and one observes the usual contribution to the expansion rate at $w_n=-1$ or $w_n=0$.\\
Finally we note that, when $M\rightarrow\infty$, some of the second order terms vanish,
as they did in the standard case $\alpha=0$. 
In fact these terms derive from the radion dynamics: 
such dynamics is sensible to the state equation of the fluids on the branes and generates a 
complicated $w_i$ dependence in (\ref{friedeq}-\ref{acceleq}). In the limit $M\rightarrow\infty$ when 
the radion is fixed to the minimum of the stabilizing potential $r$ 
it becomes trivial and these terms vanish.
\section{Conclusions}
We have examined the four-dimensional cosmological equations deriving from the 
Einstein equations  for a brane-world with a stabilizing 
potential in the presence of a Gauss-Bonnet term in the action. 
We found solutions performing an expansion up to the second order in $\rho/V$
in order to examine the cosmological behavior on the negative tension brane. 
The formalism can be easily extended to describe the 
positive tension brane as well.\\
Due to the Gauss-Bonnet extra terms, the system admits 
two different static solutions: 
one behaves as the usual RS when the Gauss-Bonnet coupling $\alpha$ goes to zero, 
while the second one has a warping factor independent of the 
bulk content in the limit
$ \alpha \rightarrow 0$  .\\
At first order the deviations from standard (without GB term) equations can be 
reabsorbed  with a redefinition of the 4-dimensional 
Newton constant. The same holds when one considers 
second order equations with a vanishing $\rho_p$. 
On the other hand, if one considers the contributions due to 
matter perturbations on the positive tension brane some deviations appear. 
An interesting feature emerging is that in the limit $\alpha\rightarrow 0$ such deviations are not swept away
when perturbing the solution that does not reduce to the usual RS.
\par
We would like to thank R.~Casadio for the helpful comments.

\end{document}